\documentclass[%
 reprint,
superscriptaddress,
 amsmath,amssymb,
 aps,
]{revtex4-2}

\usepackage{graphicx}
\usepackage{dcolumn}
\usepackage{bm}
\usepackage{comment}%
\usepackage{amsmath}%
\usepackage{url}%
\usepackage{color}%
\usepackage{tablefootnote}%

\begin{document}

\preprint{APS/123-QED}

	\preprint{APS/123-QED}
	\title{Jamming, Yielding, and Rheology during Submerged Granular Avalanche}
	\author{Zhuan Ge}
\affiliation{ Key Laboratory of Coastal Environment and Resources of Zhejiang Province (KLaCER), School of Engineering, Westlake University, 18 Shilongshan Street, Hangzhou, Zhejiang 310024, China.\\}
\affiliation{Department of Physics, College of Mathematics and Physics, Chengdu University of Technology, Chengdu, 610059, China\\}
	\author{Teng Man}%
\affiliation{ Key Laboratory of Coastal Environment and Resources of Zhejiang Province (KLaCER), School of Engineering, Westlake University, 18 Shilongshan Street, Hangzhou, Zhejiang 310024, China.\\}
\author{Kimberly M. Hill}%
\email{kmhill@umn.edu}
\affiliation{
	Department of Civil, Environmental, and Geo-Engineering, University of Minnesota, Minneapolis, Minnesota, USA
}%
\author{Yujie Wang}
\email{yujiewang@sjtu.edu.cn}
\affiliation{Department of Physics, College of Mathematics and Physics, Chengdu University of Technology, Chengdu, 610059, China\\}
\author{Sergio Andres Galindo-Torres}
\email{s.torres@westlake.edu.cn}
\affiliation{ Key Laboratory of Coastal Environment and Resources of Zhejiang Province (KLaCER), School of Engineering, Westlake University, 18 Shilongshan Street, Hangzhou, Zhejiang 310024, China.\\}
\date{\today}

\begin{abstract}
Jamming transitions and the rheology of granular avalanches in fluids are investigated using experiments and numerical simulations. Simulations use the lattice-Boltzmann method coupled with the discrete element method, providing detailed stress and deformation data. Both simulations and experiments present a perfect match with each other in carefully conducted deposition experiments, validating the simulation method. We analyze transient rheological laws and jamming transitions using our recently introduced length-scale ratio $G$. $G$ serves as a unified metric for the pressure and shear rate capturing the dynamics of sheared fluid-granular systems. Two key transition points, $G_{Y}$ and $G_{0}$, categorize the material's state into solid-like, creeping, and fluid-like states. Yielding at $G_{Y}$ marks the transition from solid-like to creeping, while $G_{0}$ signifies the shift to the fluid-like state. The $\mu-G$ relationship converges towards the equilibrium $\mu_{eq}(G)$ after $G>G_0$ showing the critical point where the established rheological laws for steady states apply during transient conditions. 
\end{abstract}

\maketitle


Submerged granular materials are ubiquitous in natural and engineering systems, such as fresh concrete, debris flows, landslides, and particulate flows in chemical engineering and food processing \cite{sousa2019drying,cao2017numerical,seguin2011dense,1995The,2002Avalanche,yang2020pore,miller2017tsunamis}. Due to the inherent characteristics of granular materials, where Brownian motion is negligible, it is difficult to separate the grain scale from the macroscopic flow scale, and the particles are influenced by surrounding fluids \cite{andreotti2013granular}. These materials exhibit physical phenomena such as memory effects, dynamic heterogeneity, avalanches, and shear thickening.
The earth and planetary landscapes are composed of granular-fluid mixtures \cite{jerolmack2019viewing}.  
\begin{figure}
    \centering
    \includegraphics[scale=0.32]{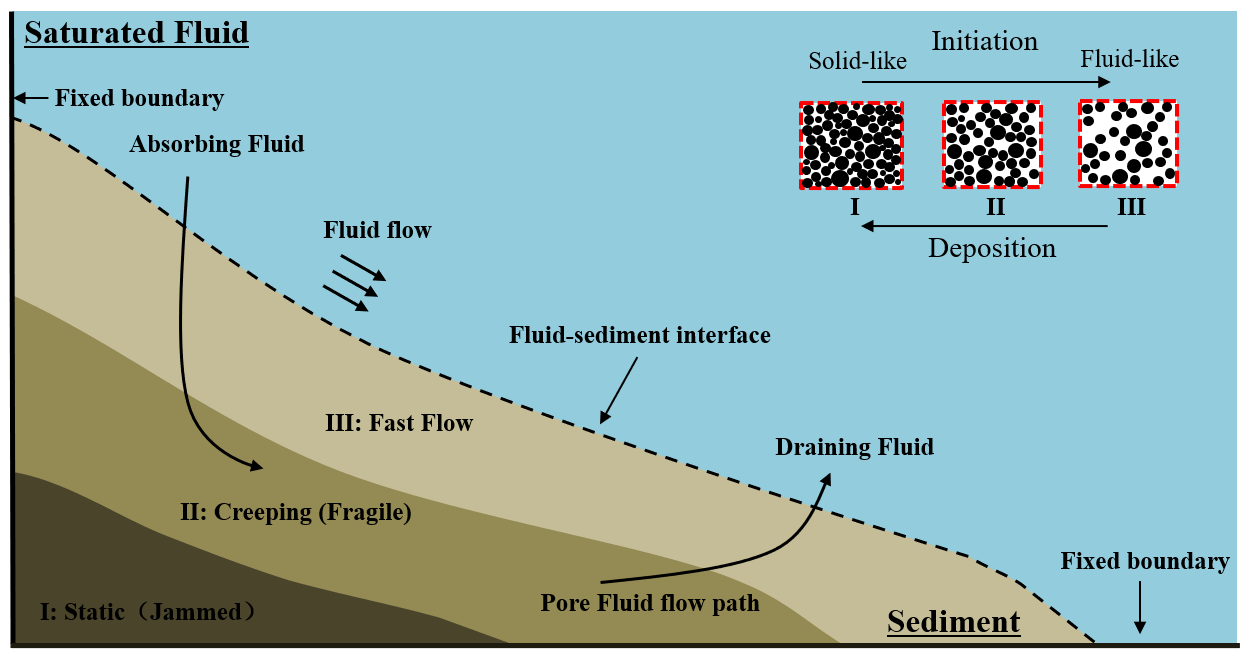}
    \caption{Sketch of a submerged avalanche showing the sediment transitioning between static, creeping, and fast flow regimes with dense, moderate, and loose solid fractions, respectively.}
    \label{fig:sub}
\end{figure}
Submerged granular avalanches are catastrophic mass movements that occur beneath the ocean surface, typically in deep-sea environments, as is shown in Fig. \ref{fig:sub}. These events involve the rapid failure and down-slope movement of sediment and rock materials along submarine slopes. Within the interface between granular material and fluid, the fluid flow reshapes the granular surface, while sediments near the fluid-granular interface exhibit fluid-like flow behavior and typically have a low solid fraction. In the middle of the granular material, sediment flow occurs in a creeping manner, with smaller deformation and motion. Granular materials near solid boundaries, such as the seafloor or main continental coastal areas, remain in a static state, exhibiting minimal deformation and a stable structure. Due to fluid-solid interaction and particle-particle collisions, sudden changes can occur among these three regimes, with the granular material transitioning from a flowing state to a creeping flow and eventually settling into a static state\cite{lacaze2021immersed}.

The dynamics and states of fluid-granular systems are influenced by confining pressure $P$, shear rate $\dot{\gamma}$, fluid viscosity $\eta_{f}$, particle density $\rho_{s}$, and particle diameter $d$. Dimensionless numbers such as the inertial number \cite{GDR2004On, jop2006constitutive} $I=\dot{\gamma}d/(\sqrt{P/\rho_{s}})$ and the viscous number \cite{cassar2005submarine, boyer2011unifying} $J=\dot{\gamma}\eta_{f}/P$ are proposed to unify the rheological relations under equilibrium states in both the inertial (where inertial effects are dominant) and viscous regimes (where fluid effects are dominant). In our previous work\cite{ge2024unifying}, we proposed a length-scale ratio-based dimensionless number $G=12(J+\lambda(St)I^{2})$ from fundamental physical analysis to unify the evolution of the solid fraction $\phi$ and the apparent friction $\mu$ from inertial to viscous regimes. Here, $St=I^{2}/J=\rho_{s}\dot{\gamma}d^{2}/\eta_{f}$ is the Stokes number, and $\lambda(St)=\frac{1-e^{-18/St}}{18-St\times(1-e^{-18/St})}$. In transient phenomenona such as granular avalanches, where the state transitions among different regimes and is influenced by deformation history, hysteresis exists regardless of whether the conditions are subaerial \cite{dijksman2011jamming,kuwano2013crossover,degiuli2017friction} or submerged\cite{perrin2019interparticle,lacaze2021immersed}. When a static granular material is subjected to applied shear stress, the magnitude of stress is required to exceed a certain threshold for the flow initiation. This characteristic, which allows the existence of meta-stable states, has the potential to give rise to catastrophic events such as earthquakes or sub-aerial landslides\cite{scholz1998earthquakes,degiuli2017friction}. The origin of hysteresis in granular flows remains a contentious topic. Nevertheless, most hypotheses proposed hitherto hinge on the existence of inertia at the particle scale\cite{degiuli2017friction,quartier2000dynamics}. This leads to a non-monotonic behavior of $\mu(\mathcal{X})$ when $\mathcal{X}\leq \mathcal{X}_{0}$ in contrast to the steady-state rheology findings, where $\mathcal{X}$ represents the governing dimensionless numbers under different conditions \cite{degiuli2017friction,perrin2019interparticle}. The hysteresis can be described by the change in frictional property for deformation paths, for instance, the apparent friction for a granular assembly to start deforming is larger than that for a system to stop moving, i.e., $\mu_{start}>\mu_{stop}$\cite{carrigy1970experiments,daerr1999two,pouliquen2002friction}. In the case of a dry granular system, where inertia is the dominant factor, $\mathcal{X}=I$\cite{degiuli2017friction}. On the other hand, in a highly viscous suspension, where viscous effects prevail, $\mathcal{X}=J$\cite{perrin2019interparticle}. In this study, we investigate the rheological laws and jamming transitions of the submerged granular avalanches using our previously proposed length-scale ratio $G$ to address two important questions: (1) Can the transitions in granular systems be unified? (2) How does granular flow behave during transient granular avalanches?

\begin{figure}
	\centering
	\includegraphics[scale=0.25]{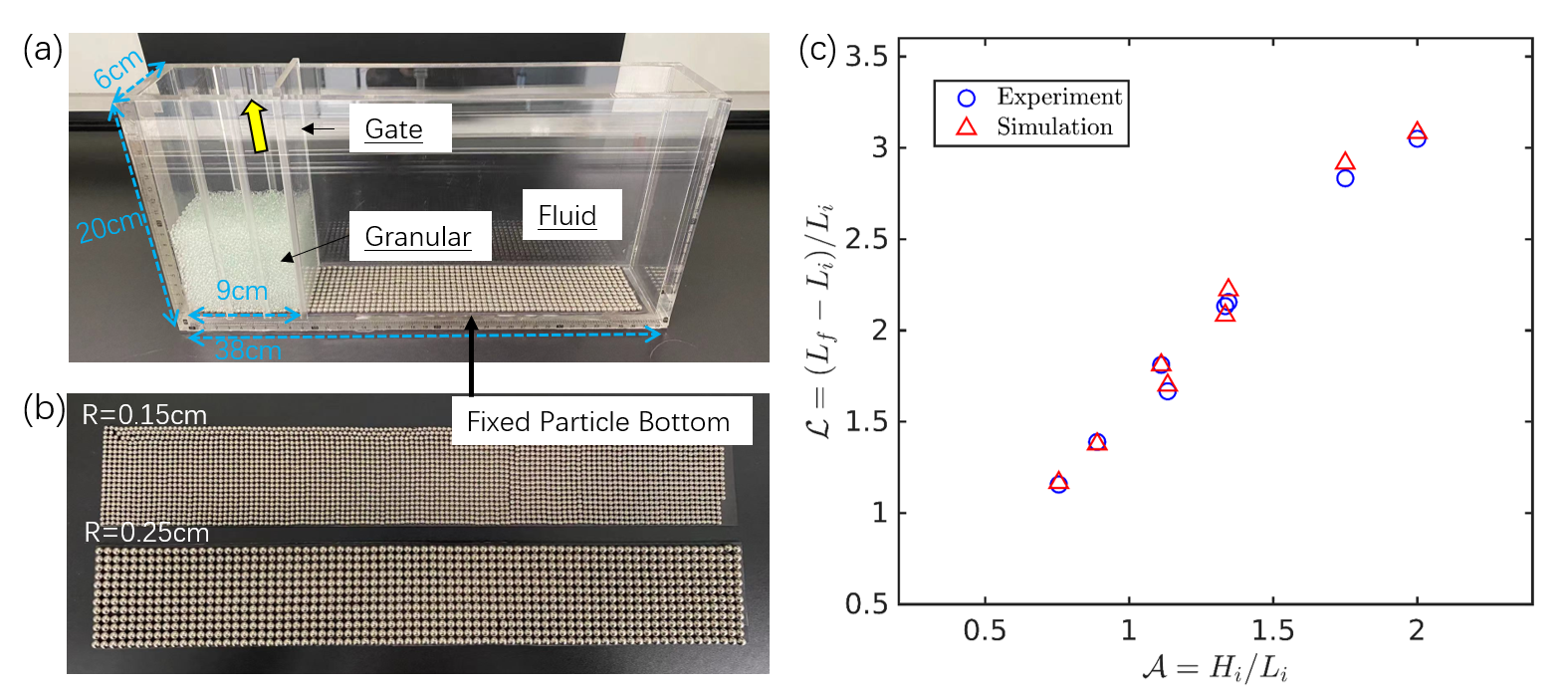}
	\caption{(a) Experimental setup of the submerged granular column collapse. (b) Dimensionless run-out length ($\mathcal{L}=(L_{f}-L_{i})/L_{i}$) as a function of the aspect ratio $\mathcal{A}=H_{i}/L_{i}$ of granular column collapse in fluid for experiments and simulation results.}
	\label{SGCsetup}
\end{figure}

As presented in Fig. \ref{SGCsetup}, the dimension of the transparent plastic tank is 38 cm$\times$6.5 cm$\times$20 cm. Three positions are considered for the vertical retaining gate, corresponding to three different initial column lengths $L_{i}=$ 3, 6, and 9 cm to generate different sizes of initial granular columns. Glass beads are used in this study. Their density is 2.5 g/cm$^3$, frictional coefficient is 0.28$\pm$0.01, and radius is 0.15$\pm$0.004 cm. Each test is recorded by a high-resolution camera with a frame rate of 100 fps. The particles are immersed in water for which the dynamic viscosity is 1 cP and the density is 1 g/cm$^3$. First, the retaining wall is placed at the desired position. Glass particles are then gently poured into the reservoir delimited by the wall to generate the initial granular column, after which we pour the liquid into the tank until it reaches the desired level. Once the fluid surface and particles are static, we measure the initial length $L_{i}$ and initial height $H_{i}$ of the granular column. Then, the retaining wall is removed suddenly, and the column collapses and propagates into the tank. When particles stop propagating, we measure the deposit length, $L_{f}$, which is the final front position, and the final peak height, $H_{f}$. In this work, the initial aspect ratio, $\mathcal{A}=H_{i}/L_{i}$, of the granular column is varied within the range of 0.75-2.

We use DEM with frictional contact interactions modeled by a Hookean contact law with energy dissipation \cite{cundall1979discrete}. LBM is used to simulate the fluid flow in the pore space and to calculate the momentum exchange between the fluid and the particles \cite{galindo2013coupled}. As shown in Fig. \ref{SGCsetup}(b), the normalized run-out distance $\mathcal{L}=(L_{f}-L_{i})/L_{i}$ shows good agreement between experiments and numerical simulations. 
\begin{figure}
	\centering
	\includegraphics[scale=0.38]{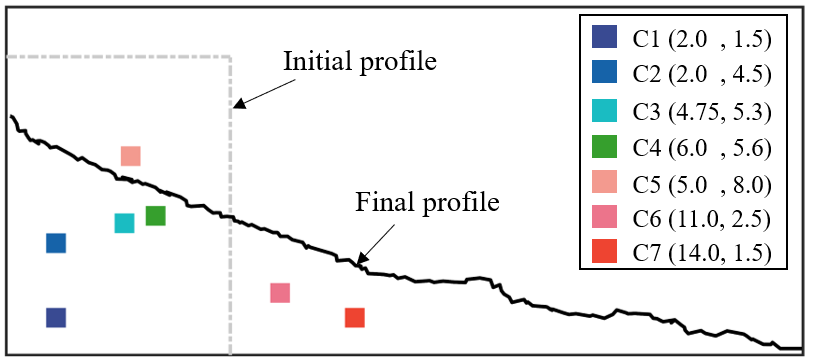}
	\caption{Seven locations are selected to track the evolution of the continuum parameter changes.}
	\label{Point}
\end{figure}
\begin{figure*}
	\centering
	\includegraphics[scale=0.5]{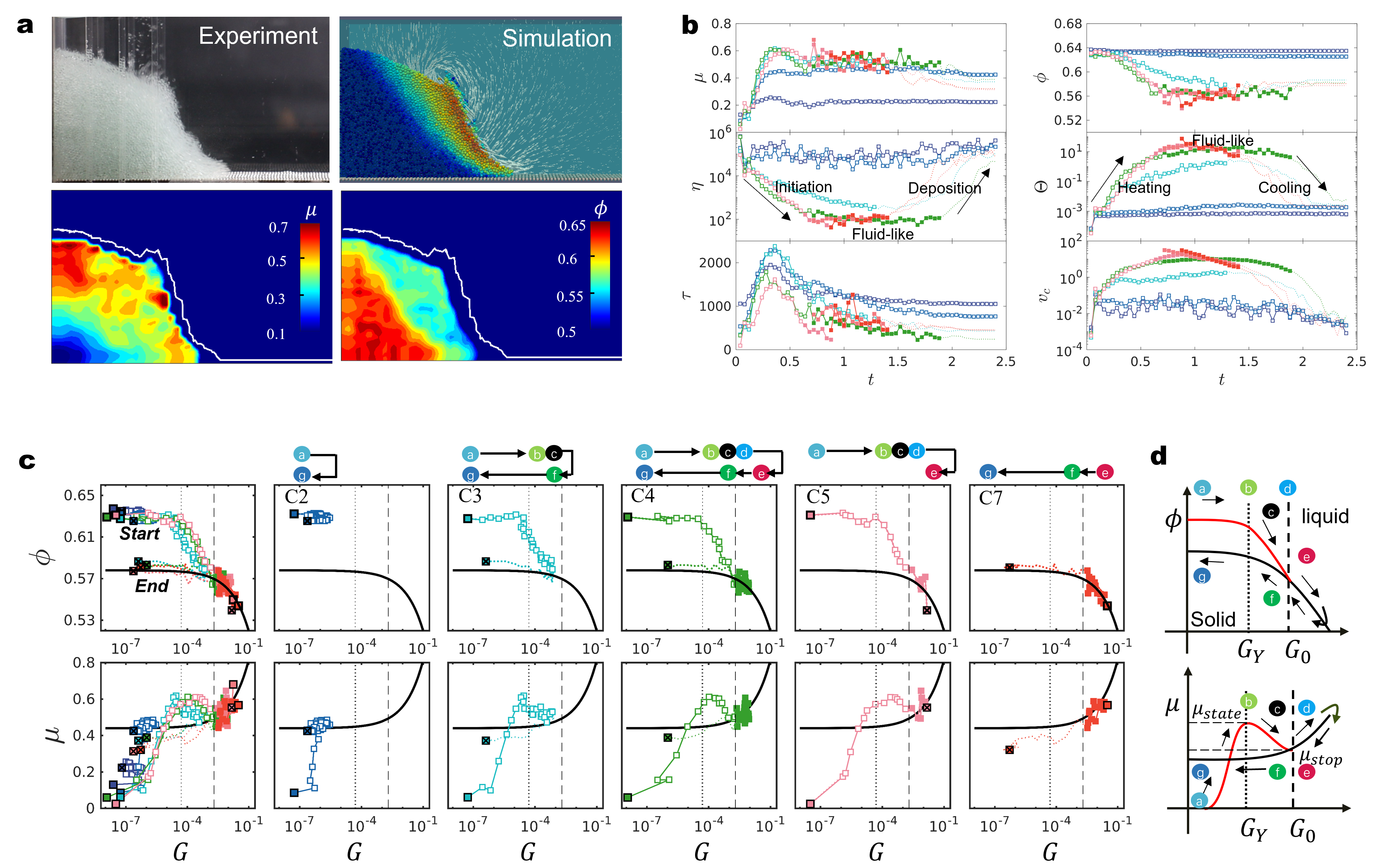}
	\caption{(a) Particle distributions of granular collapse underwater at time t=0.3 are presented for the experiment, numerical simulation, apparent friction coefficient $\mu$, and solid fraction $\phi$. (b) Evolution of the continuum parameters:  apparent frictional coefficient $\mu$, solid fraction $\phi$, viscosity $\eta$, dimensionless granular temperature $\Theta$, shear stress, $\tau$, and velocity magnitude $v_{c}$ in terms of time at the seven locations. The symbols are color-coded to correspond to the positions shown in Figure \ref{Point}: squares with a white center face represent the initiation stage, solid squares represent the fast flow stage, and dot points represent the deposition stage. (c) Plots of $\phi$and $\mu$ versus $G$ for the seven locations from initiation to deposition. Figure (d) summarize the rheological laws during the submerged granular collapse. The solid black lines in each figure are fitted by Eqs. \ref{Eq:MU} and \ref{Eq:PHI} using the data of the fast flow regime ($G\geq G_{0}$), with $\phi_{c}=0.578$, $\mu_{c}=0.44$, $a=2.7$, $b=0.32$. Different symbols represent the same data as in Figure \ref{Point}. Solid squares with black edges represent the initial step, while solid squares with a cross symbol represent the final step. Symbols of spheres represent different stages: $\textcircled{a}$ represents the initial state; $\textcircled{b}$ represents the transition from a static state to a creeping state, where $\phi$ starts to decrease and $\mu$ reaches the maximal value; $\textcircled{c}$ represents the hysteresis process; $\textcircled{d}$ is the transition point from creeping flow to the fluid-like fast flow; $\textcircled{e}$ represents the fast flow stage; $\textcircled{f}$ is the deposition stage; $\textcircled{g}$ is the final static stage. }
	\label{GPHIMUSTEP}
\end{figure*}

We implement immersed granular column collapse simulations at the initial aspect ratio $\mathcal{A}$=1.33 (the initial height is 12 cm and the initial length is 9 cm) considering various viscosities ($\eta_{f}=0.5$ g/(cm$\cdot$s), $\eta_{f}=0.1$ g/(cm$\cdot$s), $\eta_{f}=0.01$ g/(cm$\cdot$s)) and initial packing densities (dense and loose). The granular system is then discretized into several representative volume elements (RVE) with side 0.9 cm, the particles in the surrounding four cubics of each cell are used to obtain the macroscopic information such as the averaged stress $\langle\sigma \rangle$, strain rate $\dot{\gamma}$, solid fraction $\phi$, and granular temperature $T=(\sum_{p\in V}\sum_{i=x,y,z}{(\delta v_{i}^{p})}^2/D)$. $\delta v_{i}^{p}= v_{i}^{p}-v_{i}^{c}$ is the velocity fluctuation, where $v_{i}^{c}=\sum_{p\in V}v_{i}^{p}/N^{p}$ is the averaged velocity, $N^{p}$ is the number of particles, $V$ is the volume of the RVE, $v_{i}^{p}$ is the particle velocity and $D$ is the space dimension. The solid fraction in each cell is calculated by $\phi=\sum_{p\in V}V^{p}/V$, where $V^{p}$ is the volume of particle $p$. The averaged stress is calculated by the contributions of the contact term and the kinetic fluctuation $\langle \sigma \rangle=\sigma^{c}_{ij}+\sigma^{K}_{ij}=\frac{1}{V}\sum_{p\in V}f_{i}l_{j}+\frac{1}{V}\sum_{p\in V}m^{p}\delta v_{i}^{p}\delta v_{j}^{p}$, where $f_{i}$ is the $i$ component of the contact force between colliding DEM particles, and $l_{j}$ is the $j$ component of the branch vector, and $i,j$ represents the $x, y, z$ direction. The pressure $P$ and the shear stress $\tau$ are given by $P=1/3\langle\sigma_{ii}\rangle$ and $\tau=\sqrt{1/2\tau_{ij}\tau_{ij}}$, respectively, where $\tau_{ij}=\langle\sigma_{ij}\rangle-P_{p} \delta_{ij}$ is the deviatoric stress tensor. The equivalent strain rate tensor $\langle \dot{\gamma}_{ij}\rangle$ is calculated using the coarse-grain approach as described in \cite{goldhirsch2002microscopic}. As introduced in \cite{kim2020power}, the dimensionless granular temperature $\Theta=T\rho_{s}/P$ is used to understand the dynamic of the granular flow. The viscosity of each RVE is $\eta=\tau/\dot{\gamma}$, where $\dot{\gamma}=\sqrt{\dot{\gamma}^{c}_{ij}\dot{\gamma}^{c}_{ij}}$, $\dot{\gamma}^{c}_{ij}=\langle \dot{\gamma}_{ij}\rangle-1/3\dot{\gamma}^{c}_{ii}\delta_{ij}$. 

As the boundary condition changes (where the gate was removed), the original static state of the granular column is broken, and the granular assembly starts to flow. This instability transmits to the inside of the column, which leads to the granular transition from a static state to a fluid state, while some material remains at rest. 
In order to gain a better understanding of the collective behavior of the granular during the collapse process, we present the evolution of various continuum parameters, such as the solid fraction, apparent friction, strain rate, and viscosity, at six representative locations (as shown in Fig.\ref{Point}). Two of these locations (C1 and C2) are close to the bottom and wall, far away from the flow interface. At the start of the collapse, the granular at locations C3, C4, and C5 will transition from a solid state to a flow state, and eventually come to a stop. Location C6 had no grains at the beginning, and the material will flow across it during the collapse.
As seen in Fig.\ref{GPHIMUSTEP}(b), the average speed of C3, C4, and C5 increased from a low to a high value, indicating that the granular material shifted from a static state to a flowing one and eventually returned to a static state. On the other hand, C1 and C2 stayed at a low velocity, suggesting that they remained in a static state. During the initiation process, the granular assemblies were exposed to a heating process from a static state to a fast flow state, as shown in Fig.\ref{GPHIMUSTEP}(b), and their viscosity decreased. However, in the deposition process, the granular assemblies experienced a cooling process from a fast flow states to a static state, and their viscosity increased.

The rheological relations during submerged granular collapse are studied by comparing them with the $G$-based rheology obtained in the steady state\cite{ge2024unifying}. We plot $\mu$ and $\phi$ against $G$ as shown in Fig. \ref{GPHIMUSTEP}(c). A non-monotonic behavior of $\mu-G$ is observed for small $G$, specifically when $G\leq G_{0}\approx$0.002. During the initiation process, $\phi$ starts decreasing from the dense initial packing when $G$ surpasses $G_{Y}$ which is the value signaling this transition, evolving to a looser packing as $G$ increases from the beginning to the flow state. Meanwhile, $\mu$ increases to a maximal critical value ($\mu_{start}$, indicating the start of granular assembly deformation) as $G$ exceeds $G_{Y}$. A hysteresis effect occurs, similar to findings in dry granular systems \cite{degiuli2017friction} and fluid-granular systems at the viscous regime \cite{perrin2019interparticle}.

\begin{figure*}
	\centering
	\includegraphics[scale=0.72]{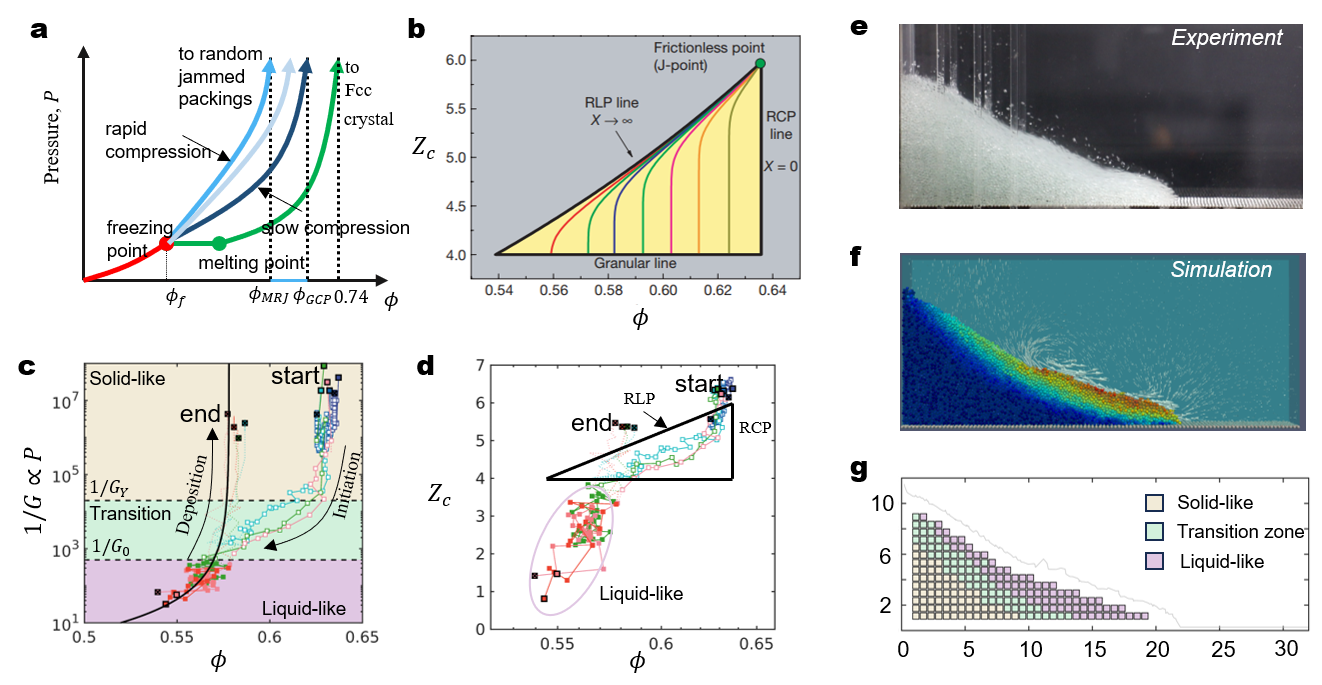}
	\caption{(a) Phase diagram in the $P-\phi$ plane for hard spheres \cite{torquato2010jammed,parisi2010mean}. (b) $Z_{c}-\phi$ phase diagram for granular matter\cite{song2008phase}. (c) $1/G-\phi$ relation, $1/G=P\cdot f(\dot{\gamma},d,\rho_{s},\eta_{f})$ (d) $Z_{c}-\phi$ relation of the granular collapse process from the initial state to the final state. Snapshot of submerged granular column collapse:(e) experiment (f) simulation at time $t=0.8$s with a dense initial packing and viscosity of fluid is 1 cP. (g) State of granular separated according to the $1/G-\phi$ phase diagram for the simulation.}.
	\label{Fig:Jammed}
\end{figure*}
In the Mohr-Coulomb failure criterion, when a static granular material is subjected to an imposed shear stress, the stress is required to exceed the critical stress (or the apparent friction is required to exceed a critical value) to initiate the flow, otherwise, the flow ceases. As the granular material begins to deform, $\mu$ decreases with increasing $G$ and becomes a growing function of $G$ when $G>G_{0}$. It's worth noting that $\mu-I$ and $\mu-J$ also present similar trend as $\mu-G$. However, the transition points at which hysteresis disappears vary across different flow regimes, whereas for $G$ they are invariant, as is shown in the supplementary material. During the deposition process, the granular material transitions from a flowing state to a solid-like state. The solid fraction $\phi$ increases to reach the final jammed packing density, while the frictional coefficient decreases to a final value.

We found, as $G\leq G_{Y}$, the granular assembly stays at the solid-like state as in C1 and C2, $\phi$ does not change and $\mu<\mu_{start}$. When $G> G_{Y}$, the granular material creeps with a small velocity as in C3. When $G> G_{0}\approx$ 0.002, the granular material presents a fast flow state (C4-C7). Furthermore, we found that $G_{0}$ and $G_{Y}$ is constant among different flow regimes and with different initial packing densities as is shown in the supplementary material. Hence we use the following formula to describe the rheological laws at the fast flow regime.
\begin{subequations}
\begin{equation}
    \mu=\mu_{c}(1+aG^{0.5}), \rm{and} \label{Eq:MU}
\end{equation}
\begin{equation}
    \phi=\phi_{c}(1-bG^{0.5}).
    \label{Eq:PHI}
\end{equation}
\end{subequations}
The fit parameters $a=$2.7 and $b=$0.32.

From a physics point of view, the state of local granular materials during a granular avalanche can be understood through the concept of jamming transition. This concept helps illuminate the rich and complicated dynamics of the process, establishing a fundamental and universal theory for predicting granular avalanches. It is crucial to propose a physically based model that indicates the jamming transition between fluid-like and disordered solid-like states, as it directly influences the physical processes involved in establishing constitutive laws to reproduce or predict real granular avalanches. In the following discussion, we will demonstrate that the transient submerged granular avalanche process bears a high similarity to the phase transition observed in ideal liquid, hard-sphere systems\cite{parisi2010mean}, and steady granular systems.

The hard-sphere phase behavior is depicted in Fig. \ref{Fig:Jammed}(a). As densities in the range of zero to the ``freezing'' point ($\phi=0.49$), the thermodynamically stable phase is a liquid. A very slow compression to increase the density leads to a first-order phase transition from liquid to a crystal branch that begins at the melting point which is the maximally dense face-centered cubic (fcc) packing ($\phi \approx 0.74$) where each particle contacts 12 others.
However, compressing a hard-sphere liquid rapidly, while suppressing significant crystal nucleation, can produce a range of meta-stable branches whose density endpoints are randomly jammed packings, akin to glasses. A rapid compression leads to a lower random jammed density compared to a slow compression. The most rapid compression presumably leads to the maximally random jammed (MRJ) state with $\phi=0.64$. As discussed in \citet{parisi2010mean}, an ideal glass state ($\phi_{GCP}\approx0.68$) exists between MRJ and fcc. Unfortunately, there is no unique meta-stable branch.

In our fluid-granular system, which accounts for friction between particles and involves transient processes commonly found in nature, the inverse of dimensionless number $1/G(P,\dot{\gamma},d,\rho_{s},\eta_{f})=P f(\dot{\gamma},d,\rho_{s},\eta_{f})$ can be interpreted as a re-scaled pressure influenced by the shear rate and material properties, where $f(\dot{\gamma},d,\rho_{s},\eta_{f})=1/(12\eta_{f}\dot{\gamma}+12\lambda(St)\dot{\gamma}^{2}d\rho_{s})$. Hence, we plot the $\phi-1/G$ relationship in Fig. \ref{Fig:Jammed}(c) to illustrate the phase transitions between fluid-like and solid-like states during transient granular collapse across all spatio-temporal domains, replicating the behavior observed in hard-sphere phase transitions. The two paths shown in Fig. \ref{Fig:Jammed}(c) resemble the two rapid compression paths in Fig. \ref{Fig:Jammed}(a), particularly considering our maximum packing density starts from $\phi\approx0.64$. This suggests that the phase transitions occurring in both the initiation stage (from solid-like to liquid-like) and deposition stage (from liquid-like to solid-like) are second-order phase transitions. With the two transition points $1/G_{0}$ and $1/G_{Y}$, the state of granular during avalanche can be separated into liquid-like, transition zone, and solid-like state as is shown in Fig.\ref{Fig:Jammed}(g).

Finally, we calculate the mechanical coordinate number $Z_{c}$, which plays a crucial role in characterizing the mechanical behavior of granular materials\cite{behringer2018physics,song2008phase}. $Z_{c}$ is calculated by the averaged contacts with no-zero force per particle in each granular assembly. The concept of $Z_{c}$ originates from the analysis of the packing structure and inter-particle forces in granular systems, providing valuable insights into their macroscopic behavior. In granular systems, particles interact through contact forces, and the mechanical coordination number provides a measure of the connectivity or degree of contact between particles. A higher mechanical coordination number indicates a denser packing of particles with more contacts, which can lead to increased structural stability and resistance to deformation. Understanding the mechanical coordinate number $Z_{c}$ is essential for predicting the response of granular materials under various loading conditions, such as compression, shear, or impact. \citet{song2008phase} conducted a great analytical work to deduced the relationship between $Z_{c}$ and $\phi$, proposed the $Z_{c}-\phi$ phase diagram to predicate the jammed state of granular materials as shown in Fig. \ref{Fig:Jammed}(b). The jammed state of granular are predicted located in the yellow regime, and $\phi$ of the jammed state with a lower bound $\phi=0.54$, a upper bound $\phi=0.63$.
We found that the fast flow regime identified by $G$ (where may be called liquid like regime as the description of hard-sphere pressure-packing fraction plane) surprisingly located out of the jammed region defined by the analytical results proposed by \citet{song2008phase}, as is shown in Fig. \ref{Fig:Jammed}(d). This finding supports and validates the description of the transition of granular materials during collapse from the perspective of jamming physics theory. Moreover, the ultimate configurations of granular assemblies consistently approximate a random loose packing (RLP), and their deposition paths, as depicted in the $1/G-\phi$ diagram, for those assemblies that have undergone a liquid-like phase, closely adhere to the rheological law governing steady-state flow (Eq. \ref{Eq:PHI}), where $\phi$ converges towards the critical state $\phi_{c}$. The studies presented in \cite{sollich1997rheology,falk2011deformation,haxton2007activated} conclude that the steady flow state of glassy materials can be interpreted as a marginal state existing at the boundary between solid and liquid phases. This theoretical framework is also applicable to the frictional fluid-granular system, as evidenced by the findings of the present work.

To summarize, the entire granular collapse process can be divided into three states based on the $1/G-\phi$ phase diagram with two transition points, $G_{0}$ and $G_{Y}$. In the solid-like state, $G\leq G_{Y}$. The transition zone occurs where $G_{Y}<G\leq G_{0}$. Finally, the fluid-like state is where $G > G_{0}$. For spatially and temporally discretized units throughout the entire granular collapse process, the fluid-like state defined by $1/G-\phi$ phase diagram closely aligns with previously proposed theoretical solutions for the phase diagram $Z_{c}-\phi$. This finding supports and validates the description of the transition of granular materials during collapse from the perspective of jamming physics theory. Additionally, the predictability of the $1/G-\phi$ phase diagram is enhanced, as $G$ is an input property of the materials and implementation conditions, rather than a response variable like $Z_{c}$. This work suggests $1/G-\phi$ relation as an alternative option of the state equation for the granular system. 

\bibliography{apssamp}

\end{document}